\documentclass[12pt,a4paper]{article}
\usepackage{epsfig}

\newcommand{\<}{\langle}
\renewcommand{\>}{\rangle}

\newcommand{\beq}{\begin{equation}}
\newcommand{\eeq}{\end{equation}}
\newcommand{\beqn}{\begin{eqnarray}}
\newcommand{\eeqn}{\end{eqnarray}}
\newcommand{\nn}{\nonumber}

\usepackage{chir_layout}
\begin{document}
\begin{titlepage}
\date{}
\vspace{-2.0cm}
\title{
{\vspace*{-6mm}
}%[-12mm]
{Reducing cutoff effects in maximally twisted LQCD 
close to the chiral limit}
\vspace*{10mm}}
\author{R.\ Frezzotti$^{a}$,
        G.\ Martinelli$^{b}$,
        M.\ Papinutto$^{c}$,
        G.C.\ Rossi$^{d,c}$
        \\[5mm]
  {\small $^a$ INFN, Sezione di Milano} \\
  {\small and Dipartimento di Fisica, Universit\`a di Milano 
   ``{\it Bicocca}''}\\
  {\small Piazza della Scienza 3 - 20126 Milano, Italy}\\
  {\small $^b$ Dipartimento di Fisica, Universit\`a di Roma
   ``{\it La Sapienza}''}\\
  {\small and INFN, Sezione di Roma ``{\it La Sapienza}''}\\
  {\small Piazzale A. Moro - 00185 Roma, Italy}\\
  {\small $^c$ John von Neumann-Institut f\"ur Computing NIC}\\
  {\small Platanenallee 6, D-15738 Zeuthen, Germany}\\
  {\small $^d$ Dipartimento di Fisica, Universit\`a di Roma
   ``{\it Tor Vergata}''}\\
  {\small and INFN, Sezione di Roma 2}\\
  {\small Via della Ricerca Scientifica - 00133 Roma, Italy}}

%%%%%%%%%%%%%%%%%%%%%%%%%%%%%%%%%%%%%%%%%%%%%%%%%%%%%%%%%%%%%%%%%%%%%%%%

\maketitle
\vspace*{0mm}
\abstract
{When analyzed in terms of the Symanzik expansion, lattice correlators 
of multi-local (gauge-invariant) operators with non-trivial continuum limit 
exhibit in maximally twisted lattice QCD ``infrared divergent'' cutoff 
effects of the type $a^{2k}/(m_\pi^2)^{h}$, $2k\geq h\geq 1$ ($k,h$ integers), 
which tend to become numerically large when the pion mass gets small. We prove 
that, if the action is O($a$) improved {\it \`a la} Symanzik or, alternatively, 
the critical mass counter-term is chosen in some ``optimal'' way, these 
lattice artifacts are reduced to terms that are at worst of the order 
$a^{2}(a^2/m_\pi^2)^{k-1}$, $k\geq 1$. 
This implies that the continuum extrapolation of lattice results is smooth 
at least down to values of the quark mass, $m_q$, satisfying the order of 
magnitude inequality $m_q >a^2\Lambda^3_{\rm QCD}$.}
\end{titlepage}

\section{Introduction and main results}
\label{sec:INTRO}

Although in lattice QCD at maximal twist (Mtm-LQCD) O($a$) discretization 
effects (actually all O($a^{2k+1}$), $k\geq 0$ effects) are absent or easily 
eliminated~\cite{TM,FR1,FR2}, it turns out that, because of the explicit 
breaking of parity and iso-spin symmetry present in the action, correlators 
are affected by dangerous artifacts of relative order $a^{2k}$, $k\geq 1$, 
which are enhanced by inverse powers of the (squared) pion mass, as the latter
becomes small. Terms of this kind may have been seen in the simulations  
reported in ref.~\cite{KSUW}.

When analyzed in terms of the Symanzik expansion, the expectation 
values of multi-local (gauge-invariant) operators, $\<O\>|^L_{m_q}$, 
exhibit, as $m_\pi^2\to 0$, what we will call ``infrared (IR) divergent'' 
cutoff effects with a behaviour of the form~\footnote{Often for short we do 
not include powers of $\Lambda_{\rm QCD}$ required to match physical 
dimensions.}
\beq
\<O\>\Big{|}^L_{m_q}=\<O\>\Big{|}^{\rm cont}_{m_q}
\Big{[}1+{\rm O}\Big{(}\frac{a^{2k}}{(m_\pi^2)^{h}}\Big{)}\Big{]}
\, ,\quad 2k\geq h\geq 1 \,\,(k,h \,\,{\rm integers})\, ,\label{ORD}\eeq 
where we have assumed that the lattice correlator has 
a non-trivial continuum limit. 

It is important to remark that the existence of pion poles in the Symanzik 
expansion of a lattice correlator in no way means that the latter diverges 
as $m_q\to 0$, but only that the Symanzik expansion may break down if the 
order of magnitude inequality $m_\pi^2>a$ is violated~\cite{SH05}.

We will prove that, if  the action is O($a$) improved {\it \`a la} Symanzik 
by introducing the lattice clover term~\cite{SW} (with its non-perturbatively 
determined $c_{SW}$ coefficient~\cite{LU}), or the critical mass counter-term 
is chosen in some ``optimal'' way, these artifacts will be reduced to 
terms that are at worst of order $a^{2}(a^2/m_\pi^2)^{k-1}$, $k\geq 1$. 

The idea that a suitable definition of critical mass exists which 
can lead to a smoothing out of chirally enhanced lattice artifacts or
perhaps be of help in getting improvement was already put forward in the 
context of chiral perturbation theory ($\chi$PT) in refs.~\cite{SHWUNEW} 
and~\cite{AB}, respectively.

A crucial consequence of the analysis we present in this paper 
is that the strong (order of magnitude) inequality
\beq
m_q> a\Lambda^2_{\rm QCD}\, ,
\label{STINEQ}\eeq
invoked in ref.~\cite{FR1} in order to have the phase of the chiral 
vacuum driven by the quark mass term and not by the (twisted) Wilson 
term, can be relaxed to the weaker relation 
\beq
m_q> a^2\Lambda^3_{\rm QCD}\, ,
\label{WINEQ}\eeq
before large cutoff effects are possibly met when the quark mass is 
lowered at fixed $a$. The bound~(\ref{WINEQ}) is fairly weak as it 
allows simulations in a region of quark masses that correspond to 
rather light pions (with masses close to around 200 MeV for typical 
present-day lattice spacings).

In order to make this observation useful in practice, it is however 
necessary to have a handle on the numerical coefficients entering  
the bound~(\ref{WINEQ}). We will argue that one can monitor 
the ``safe'' region of quark masses 
by measuring the mass of the charged pion as a function of $m_q$. 
It can be shown, in fact, that, once the leading ``IR divergent'' 
cutoff effects have been eliminated (and under the assumption that chiral 
symmetry is spontaneously broken in the continuum), the lattice squared 
mass of the charged pion, for sufficiently small $m_q$, is a linear 
function of $m_q$ (with only small O($a^4$) and O($m_q^2$) distortions) 
at least until the pion mass remains within the bound hinted at by 
eq.~(\ref{WINEQ}).

The situation one finds in this kinematic regime is thus reminiscent 
of the ideal continuum case, where the Gell-Mann--Oakes--Renner (GMOR) 
relation~\cite{GMOR} ensures the proportionality of $m_\pi^2$ to the mass 
of the quark. Armed with the lattice analog of the GMOR relation one can exploit 
the previous considerations to argue that, vice-versa, at fixed $a$, $m_q$ 
should not be decreased beyond the point where non-linearities show up.

The effectiveness of Mtm-LQCD in removing O($a$) discretization errors was 
successfully tested in quenched simulations in refs.~\cite{XLF}. The 
ability of the optimal choice of the critical mass in diminishing the 
magnitude of lattice artifacts at small quark mass has been beautifully 
demonstrated in the recent works of refs.~\cite{CAN,XLFNEW,XLFSCAL}. 
Results of unquenched simulations~\cite{KARETAL}, though encouraging, 
are still rather preliminary due to the complicated phase structure~\cite{PS} 
of Wilson fermions when $m_q / \Lambda_{\rm QCD}$ is numerically
comparable to (or smaller than) $a^2 \Lambda_{\rm QCD}^2$.

The plan of the paper is as follows. In Section~\ref{sec:SEOLC} we analyze 
the form of the Symanzik expansion of lattice correlators beyond O($a$)
and explain why and how ``IR divergent'' cutoff effects arise in this context.  
In Section~\ref{sec:KLL} we discuss two ways of removing all the leading 
``IR divergent'' cutoff effects and we describe the structure of the 
left-over ``IR divergent'' terms. In Section~\ref{sec:GOR} we derive  
the lattice GMOR relation and we discuss how it can be used to monitor 
the lowest value of the quark mass that can be safely employed if the 
lattice spacing is held fixed. Finally in Section~\ref{sec:ART} we 
collect some remarks on the peculiar structure of the lattice artifacts 
affecting lattice hadron energies and the pion decay constant. 
Conclusions can be found in Section~\ref{sec:CONC}. In an Appendix we prove 
that automatic O($a^{2k+1}$) improvement of the lattice expectation values 
of parity-even operators holds in Mtm-LQCD, by using an argument which does 
not rely on the spurionic transformation $r\to -r$, hence on the 
$r$-parity properties of the critical mass.

\section{Symanzik analysis of ``IR divergent'' cutoff artifacts}
\label{sec:SEOLC}  

The expression of fermionic action of  Mtm-LQCD is given in the 
physical quark basis~\cite{FR1,FR2} by the formula
\beq
S_{\rm tmF}^L=a^4\sum_x\bar\psi^L(x)\Big{[}\gamma\!\cdot\!\widetilde{\nabla} 
+m_q-i\gamma_5\tau_3 \Big{(}-a\frac{r}{2}\nabla^*\!\cdot\!\nabla+
M_{\rm cr}^{e}(r) \Big{)}\Big{]} \psi^L(x) \, , \label{MTMA}
\eeq
where $\psi^L$ is a lattice fermion doublet, $r$ is 
the Wilson parameter and $m_q$ is the bare quark mass, 
i.e.\ the real parameter which in the continuum will provide a 
non-vanishing mass to the pion. $M_{\rm cr}^{e}$ is any sensible 
(from the point of view of renormalization theory) 
``estimate'' of the critical mass which, in order to match the $r$-parity 
properties of the Wilson term, should be taken as an odd 
function of $r$~\cite{FR1}. 

The study of cutoff artifacts affecting lattice correlators in Mtm-LQCD 
can be elegantly made in the language of the Symanzik expansion. A full 
analysis of cutoff effects beyond O($a$) is of course extremely complicated. 
Fortunately it is not necessary, if we limit the discussion to the terms 
that are most strongly enhanced as the quark mass is decreased. This 
analysis will be carried out in the next section.

For completeness we show in an Appendix that automatic O($a$)
(actually O($a^{2k+1}$), $k\geq 0$) improvement directly follows from
the symmetry of the lattice theory under the transformation 
$P\times {\cal{D}}_{d}\times (m_q\to-m_q)$, where $P$ is parity 
and the transformation ${\cal{D}}_{d}\times (m_q\to-m_q)$ counts the parity 
of the overall dimension of an operator. The explicit definition 
of the parity operation $P$ and the transformation ${\cal{D}}_{d}$  
can be found in eqs.~(\ref{PAROP}) and~(\ref{FIELDT}). 
The proof we give does not involve any change of sign of the Wilson 
parameter, thus it is independent of the $r$-parity properties of the 
critical mass~\footnote{We wish to thank M. L\"uscher for a stimulating 
discussion on this issue during the ``Twisted Mass Lattice Fermions'' 
workshop held in Villa Mondragone (Frascati - Italy) on March 14-15, 2005.}, 
unlike the argument developed in ref.~\cite{FR1}.

\subsection{The Symanzik LEEA of Mtm-LQCD}
\label{sec:SEMQCDA}

With reference to the fermionic lattice action~(\ref{MTMA}), the 
low energy effective action (LEEA) of Mtm-LQCD, $S_{\rm Sym}$,
can be conveniently written in the form
\beq
S_{\rm Sym}=\int\!d^4y\,\Big{[}{\cal L}_4(y)+
\sum_{k=0}^{\infty}a^{2k+1}\ell_{4+2k+1}(y)
+\sum_{k=1}^{\infty}a^{2k}\ell_{4+2k}(y)\Big{]}\, ,
\label{SLEEA}\eeq 
where ${\cal L}_4=\frac{1}{2g_0^2}{\rm tr}(F\!\cdot\! F)+ 
\bar\psi(\gamma \!\cdot\! D + m_q)\,\psi$ is the target 
continuum QCD Lagrangian density. A number of interesting properties 
enjoyed by the above LEEA can be proved which we summarize below.

$\bullet$ Lagrangian density terms of even dimension, $\ell_{2k}$, in 
eq.~(\ref{SLEEA}) are parity-even, while terms of odd dimension, 
$\ell_{2k+1}$, are parity-odd and twisted in iso-spin space. 
Thus the latter have the quantum numbers of the neutral pion. 
These parity properties follow from dimensional arguments and the 
invariance of the LEEA of Mtm-LQCD under the transformation 
$P\times{\cal D}_d\times(m_q\to -m_q)$, inherited from the 
correspondent invariance of the lattice theory (see Appendix).

$\bullet$ The term of order $a$ in eq.~(\ref{SLEEA}), $\ell_5$,
is given (after use of the equations of motion of continuum
QCD) by the linear combination
\begin{eqnarray}
\hspace{-1.5cm}&&\ell_5=\delta_{5,SW}\,\ell_{5,SW}+\delta_{5,m^2}\,
\ell_{5,m^2}+\delta_{5,e}\,\ell_{5,e}\, ,\label{L5}\\
\hspace{-1.5cm}&&\ell_{5,SW}=
\frac{i}{4}\bar\psi[\sigma\cdot F]i\gamma_5\tau_3\psi \, ,
\!\!\quad\ell_{5,m^2} = m_q^2 \bar\psi i\gamma_5\tau_3\psi \, ,\!\!\quad
\ell_{5,e} = \Lambda_{\rm QCD}^2 \bar\psi i\gamma_5\tau_3\psi  \, ,
\label{L51}
\end{eqnarray}
where the coefficients $\delta_{5,SW}$, $\delta_{5,m^2}$ and $\delta_{5,e}$ 
are O(1) dimensionless quantities, odd in $r$. The operator $\ell_{5,e}$ 
arises from the need to describe order $a$ uncertainties entering any 
non-perturbative determination of the critical mass and goes together 
with $\ell_{5,SW}$. Both $\ell_{5,SW}$ and $\ell_{5,e}$ could be made to 
disappear from eq.~(\ref{SLEEA}) by introducing in the Mtm-LQCD action the 
SW (clover)-term with the appropriate non-perturbatively determined $c_{SW}$ 
coefficient and at the same time setting the critical mass to its 
correspondingly O($a$) improved value. 

$\bullet$ Higher order ambiguities ($k\geq 1$) in the critical mass will be 
described by terms proportional to odd powers of $a$, more precisely of the kind 
\beq a^{2k+1}\,\delta_{4+2k+1,e}\,\ell_{4+2k+1,e}=a^{2k+1}\,\delta_{4+2k+1,e}\,
(\Lambda_{\rm QCD})^{2k+2}\,\bar\psi i\gamma_5\tau_3\psi\, .\label{HK}\eeq
The structure of these terms, which will all contribute to 
${\cal L}_{\rm odd}$ (see eq.~(\ref{DEFEO}) below), follows again from 
obvious dimensional arguments plus the invariance of the lattice action 
under the transformation $P\times{\cal D}_d\times(m_q\to -m_q)$.

\subsection{Describing Mtm-LQCD correlators beyond O($a$)}
\label{sec:SEMQCDB}

We are interested in the Symanzik description of the lattice artifacts 
affecting connected expectation values of $n$-point, multi-local, 
multiplicative renormalizable (m.r.) and gauge-invariant operators
\beq
O(x_1,x_2,\ldots,x_n)=\prod_{j=1}^{n} O_j(x_j)\equiv O(x)\, ,
\quad x_1\neq x_2 \neq \ldots \neq x_n\, ,\label{OMUL}
\eeq
which we take to have continuum vacuum quantum numbers, so as to yield 
a result that does not trivially vanish as $a\to 0$. In particular, in order 
to ensure automatic O($a$) improvement~\cite{FR1} we will assume that $O$ 
is parity invariant in which case its Symanzik expansion 
will contain only even powers of $a$. Schematically we write 
\beqn
\hspace{-.3cm}\<O({x})\>\Big{|}_{m_q}^{L}\!=\!
\<[O({x})+\Delta_{\rm odd}O(x)+\Delta_{\rm even}O(x)]
e^{-\int\!d^4 y[{\cal L}_{\rm odd}(y)+{\cal L}_{\rm even}(y)]}\>
\Big{|}^{\rm cont}_{m_q}\, , \label{SEOP}
\eeqn
where for short we have introduced the compact notations
\beq
{\cal L}_{\rm odd}=\sum_{k=0}^{\infty}a^{2k+1}\ell_{4+2k+1}\, ,\qquad
{\cal L}_{\rm even}=\sum_{k=1}^{\infty}a^{2k}\ell_{4+2k}\, .
\label{DEFEO}
\eeq
The operators $\Delta_{\rm odd}O$ (resp.\ $\Delta_{\rm even}O$)
have their origin in the need of regularizing terms where a parity-odd
(resp.\ parity-even) product of ${\cal L}_{\rm odd}$ and/or 
${\cal L}_{\rm even}$ insertions comes in contact with some of the 
points where the local operator factors appearing in $O$ are concentrated. 
$\Delta_{\rm odd}O$ (resp.\ $\Delta_{\rm even}O$) counter-terms have an 
expansion in odd (resp.\ even) powers of $a$. We recall that they can 
be viewed as the $n$-point operators necessary for the on-shell 
improvement of $O$~\cite{HMPRS,LU}.

{}For the purpose of this discussion we imagine expanding the r.h.s.\ of 
eq.~(\ref{SEOP}) in powers of $\int\!d^4y\,{\cal L}_{\rm odd}(y)$ and/or 
$\int \!d^4y\,{\cal L}_{\rm even}(y)$. 
Terms with $j$ and/or $j'$ insertions of the first and/or the second of these 
factors will generate in the Symanzik expansion $h$-fold $1/m_\pi^2$ pion 
poles with $1\leq h\leq j+j'$. 

\subsection{Pion poles and ``IR divergent'' cutoff effects}
\label{sec:PPSSE} 

Although a complete analysis of all the ``IR divergent'' cutoff effects 
is very complicated, the structure of the leading ones ($h=2k$ in 
eq.~(\ref{ORD})) is rather simple, as they only come from continuum 
correlators where $2k$ factors $\int d^4y\,{\cal L}_{\rm odd}(y)$ 
are inserted. More precisely the leading ``IR divergent'' cutoff 
effects are identified on the basis of the following 

\vskip .2cm
{\it Result:$\;$}
In the Symanzik expansion of $\<O(x)\>|^L_{m_q}$ at order $a^{2k}$ 
($k\geq 1$) there appear terms with a $2k$-fold pion pole and residues 
proportional to $|\<\Omega|{\cal L}_{\rm odd}|\pi^0({\bf 0})\>|^{2k}$, 
where $\<\Omega|$ and $|\pi^0({\bf 0})\>$ denote the vacuum 
and the one-$\pi^0$ state at zero three-momentum, respectively. 
Putting different factors together, each one of these terms can be seen 
to be schematically of the form (recall ${\cal L}_{\rm odd}={\rm O}(a)$) 
\beqn
\hspace{-1.cm}&&\Big{[}\Big(\frac{1}{m_\pi^2}\Big)^{2k}
(\xi_{\pi}(m_q))^{2k}{\cal M}[O;\{\pi^0({\bf 0})\}_{2k}]
\Big{]}_{m_q}^{\rm cont}\, ,
\label{ZLCP}\eeqn
where we have generically denoted by ${\cal M}[O;\{\pi^0({\bf 0})\}_{2k}]$ 
the $2k$-particle matrix elements of the operator $O$, with each of the 
$2k$ particles being a neutral pion at zero three-momentum and we have 
introduced the short-hand notation 
\beq
\xi_{\pi}(m_q)=\Big{|}\<\Omega|{\cal L}_{\rm odd}|
\pi^0({\bf 0})\>\Big{|}_{m_q}^{\rm cont}\, .\label{METREO}\eeq
\vskip .2cm

The proof of the above {\it Result} can be obtained on the basis of 1) the 
general theorems of quantum field theory governing the appearance 
of poles in correlators~\cite{WEIN}, 2) the notion of ``partially 
disconnected'' diagrams which come about when reducing pairs of pions 
from {\it in} and {\it out} states of a multi-pion matrix element, 3) the 
observation that the ($2k+1$)-pion matrix elements of ${\cal L}_{\rm odd}$
can all be shown to be proportional to $\xi_\pi$ in the chiral limit,
through a repeated use of soft pion theorems~\cite{SPT}, in which  
pions are successively reduced out.

Less ``IR divergent'' cutoff effects (those with $h$ strictly 
smaller than $2k$ in eq.~(\ref{ORD})) come either from terms 
with insertions of $\int d^4y{\cal L}_{\rm even}(y)$ and/or pairs of 
$\int d^4y{\cal L}_{\rm odd}(y)$ factors, or from contributions of 
more complicated intermediate states other than straight zero 
three-momentum pions or from both. In all cases one gets terms with 
extra $a^2$ powers not ``accompanied'' by an equal number of 
$1/(m_\pi^2)^2$ factors.

\section{Reducing ``IR divergent'' cutoff artifacts}
\label{sec:KLL}

We have shown in the previous sections that close to the chiral limit the
most ``IR divergent'' discretization artifacts affecting Mtm-LQCD correlators 
at order $a^{2k}$ are proportional to $2k$ powers of $\xi_\pi$ (see 
eq.~(\ref{ZLCP})). Since ${\cal{L}}_{\rm odd}=a\,\ell_5+{\rm O}(a^3)$, this 
also means that at leading order in $a$ each multiple pion pole residue is 
proportional to $|\<\Omega|\ell_{5}|\pi^0({\bf 0})\>|^{2k}$. It is an 
immediate conclusion of this analysis that all these dangerous cutoff 
effects can be removed from lattice data if we can either eliminate $\ell_{5}$ 
from the Symanzik LEEA of Mtm-LQCD or set $\xi_\pi$ to zero. 
Actually, in the last case a somewhat weaker condition is sufficient. 
Indeed, we will see that it is enough to reduce 
$\xi_\pi$ to a quantity of order $a m_\pi^2$.

\subsection{Improving the Mtm-LQCD action by the SW-term}
\label{sec:IASWT}

The obvious, field-theoretical way to eliminate $\ell_{5}$ from the Symanzik 
LEEA of Mtm-LQCD consists in making use of the O($a$) improved action 
\beqn 
&&S_{\rm tmF}^{IL}=
a^4\sum_x\bar\psi^L(x)\Big{[}\gamma\!\cdot\!\widetilde{\nabla}+ m_q +\nn\\
&&-i\gamma_5\tau_3 \Big{(}-a\frac{r}{2}\nabla^*\!\cdot\!\nabla+
M_{\rm cr}^{Ie}(r)+\frac{i}{4} c_{SW}(r) [\sigma \!\cdot\! F]^L
\Big{)} \Big{]} \psi^L(x) \, ,
\label{MTMAI}\eeqn 
where $c_{SW}$ is fixed in the appropriate non-perturbative way~\cite{LU} 
and $M_{cr}^{Ie}$ is an improved estimate of the critical mass.

In this situation
the lattice correlation functions of the theory will admit a Symanzik
description in terms of a LEEA where the operators ${\ell}_{5,SW}$ and
${\ell}_{5,e}$ are absent, and ${\ell}_5$ will be simply given by 
${\ell}_{5,m^2}$ (see eq.~(\ref{L5})). The left-over contributions 
arising from the insertions of ${\ell}_{5,m^2}$ in $\<O\>|_{m_q}^{\rm cont}$ 
will yield terms that are at most of order 
$(am_q^2/m_\pi^2)^{2k}\simeq (a m_q)^{2k}$, hence negligible in the chiral 
limit. It is instead the next odd operator in the 
Symanzik expansion, $a^3\ell_7$, which comes into play. 

A detailed combinatoric analysis based on the structure of the 
non-leading ``IR divergent'' cutoff effects reveals that the worst 
lattice artifacts left behind in correlators after the clover cure 
are of the kind $a^2(a^2/m_\pi^2)^{k-1}$, $k\geq 1$. 

\subsection{Optimal choice of the critical mass}
\label{sec:OCCM}

The alternative strategy to eliminate all the leading ``IR divergent'' cutoff 
effects consists in leaving the Mtm-LQCD action of the form~(\ref{MTMA}), 
but fixing the critical mass through the condition
\beq 
\lim_{m_q \to 0^{+}} \xi_\pi(m_q)=\lim_{m_q \to 0^{+}} \;
\Big{|}\<\Omega|{\cal{L}}_{\rm odd}|\pi^0({\bf 0})\>
\Big{|}^{\rm cont}_{m_q}\; = \; 0 \, .
\label{EFFCOND}
\eeq
The meaning of this condition is rather simple. It amounts to fix, 
for each $k\geq 0$, the order $a^{2k+1}$ contribution in the critical 
mass counter-term, $-M_{\rm cr}^{opt}\,\bar\psi^L i\gamma_5\tau_3\psi^L$, 
in such a way that its vacuum to one-pion state matrix element 
compensates, in the limit of vanishing quark mass, the similar matrix element 
of the sum of all the other operators making up $\ell_{4+2k+1}$.

In the next section we present concrete procedures designed to implement 
the condition~(\ref{EFFCOND}) in actual simulations. To avoid confusion
with the values the lattice action parameters will take in the successive
steps of a simulation, we will provisionally put a bar over the symbols 
representing the values of the quark and the corresponding pion state mass 
while we develop the argument for the ``optimal'' determination of the critical mass.

\subsubsection{Lattice estimate}
\label{sec:TP}

We want to show how eq.~(\ref{EFFCOND}), which has to do with a
matrix element defined in the continuum theory, can be translated into 
a lattice condition. To this end let us consider the lattice correlator 
\beq
a^3\sum_{\bf x}\;\<{\cal Q}(x,0)\>\Big{|}^{L}_{\bar m_q} = 
a^3\sum_{\bf x}\;\<V_0^2(x)P^1(0)\>\Big{|}^L_{\bar m_q} \, ,
\qquad x_0 \neq 0 \, ,
\label{LATCOR}\eeq
where $V_0^2=\bar\psi\gamma_0\frac{\tau_2}{2}\psi$ is the vector current 
with iso-spin index 2 and $P^1=\bar\psi \gamma_5 \frac{\tau_1}{2}\psi$ is 
the pseudo-scalar quark density with iso-spin index 1.
In the continuum the correlator~(\ref{LATCOR}) owing to parity vanishes 
for any value of $\bar m_q$, and we have 
\beq
\<Q_V^2 P^1(0)\>\Big{|}^{\rm cont}_{\bar m_q} = 0\, , \label{WTI}
\eeq
where $Q_V^2=\int\! d{\bf x}V_0^2({\bf x},t)$ is the iso-spin 2 vector 
charge. On the lattice the breaking of parity (and iso-spin) due to the 
presence of the twisted Wilson term makes the correlator~(\ref{LATCOR}) 
non-vanishing by pure discretization effects. Extending to parity violating 
correlators the arguments developed in Sect.~\ref{sec:SEMQCDA}, one gets 
for its Symanzik expansion 
\beqn
\hspace{-.8cm}&&a^3\sum_{\bf x}\;\<{\cal Q}(x,0)\>\Big{|}^{L}_{\bar m_q}=\nn\\
\hspace{-.8cm}&&=\int \!d{\bf x}\,\Big{\{}\<\Delta_{\rm odd}{\cal Q}(x,0)\>
\Big{|}_{\bar m_q}^{\rm cont}
-\<{\cal Q}(x,0)\int \!d^4y\,{\cal L}_{\rm odd}(y)\>
\Big{|}_{\bar m_q}^{\rm cont}+\ldots\, ,\label{SYMO}\eeqn
where dots represent terms with higher order insertions of 
$\int\!d^4y\,{\cal L}_{\rm odd}(y)$ and/or 
$\int\!d^4y\,{\cal L}_{\rm even}(y)$ 
and $\Delta_{\rm odd}{\cal Q}$ has the expression 
\beqn
\hspace{-.5cm}&&\Delta_{\rm odd}{\cal Q}(x,0)=\label{MIX}\\
\hspace{-.5cm}&&=a\Big{[}\eta\,\partial_{0}\bar\psi\gamma_5
\frac{\tau_1}{2}\psi(x)\,\bar\psi \gamma_5 \frac{\tau_1}{2}\psi(0)+
\tilde\eta\,\bar{m}_q\,\bar\psi\gamma_0\gamma_5\frac{\tau_1}{2}\psi(x)
\,\bar\psi \gamma_5 \frac{\tau_1}{2}\psi(0)\Big{]}+{\rm O}(a^3)\, ,\nn
\eeqn
In eq.~(\ref{MIX}) $\eta$ and $\tilde\eta$ are appropriate dimensionless 
coefficients, odd in $r$. We recall that in the language of the Symanzik 
improvement program the term we have explicitly written down in eq.~(\ref{MIX}) 
is the standard operator necessary for the on-shell O($a$) improvement of 
${\cal Q}(x,0)$. In writing the expansion~(\ref{SYMO}) use has been made 
of the continuum relation~(\ref{WTI}). 

{}For the rest of the argument it is important to remark that from 
higher order insertions of $\int\!d^4y\,{\cal L}_{\rm odd}(y)$ 
arbitrarily high powers of the ratio $\xi_\pi(\bar m_q)/\bar{m}_\pi^2$ 
will be generated in the r.h.s.\ of eq.~(\ref{SYMO}) 

At large times ($t\gg 1/\Delta m$, where $\Delta m$ is the difference
between the mass of the first excited state with the pion quantum numbers 
and that of the pion), we may write more explicitly the r.h.s.\ of 
eq.~(\ref{SYMO}) in the form
\beqn
\hspace{-1.2cm}&&\lim_{t\to+\infty}\Big{[}\frac{e^{-\bar m_\pi t}}{2\bar m_\pi}
\Big{]}^{-1}a^3\sum_{\bf x} \; \< {\cal Q}(x,0) \>\Big{|}^{L}_{\bar m_q} \; =
\; \Big{[} \, a\bar m_q\tilde\eta\,\<\Omega|A_0^1|\pi^1\!({\bf 0})\> C_{SSL} +
\nn\\\hspace{-1.2cm}&&
-a\bar m_\pi\eta\,\<\Omega|P^1|\pi^1\!({\bf 0})\> C_{SL}
-\frac{\xi_\pi(\bar{m}_q)}{2 \bar m_\pi^2} \, 2\bar m_\pi C_L \, \Big{]} \;
\<\pi^1({\bf 0})|P^1|\Omega\>\Big{|}_{\bar m_q}^{\rm cont}
\, ,   \label{SYMPI}
\eeqn
with all corrections of order $a^3$ or higher encoded 
in the coefficient factors
\beqn
\hspace{-1.8cm}&& C_X = 
1 + \sum_{j,k,\ell=1}^{\infty} c_{j,k,\ell}^{(X)} \Big{(}
\frac{\xi_\pi(\bar{m}_q)}{\bar m_\pi^2} \Big{)}^{j}
\Big{(}\frac{a^2}{{\bar m_\pi^2}}\Big{)}^{k} a^{\ell}\Big{|}_{j+\ell={\rm{even}}}\, , \quad
X=L,SL,SSL \, .
\label{SYMPICX}
\eeqn
Several observations about the r.h.s.\ of eq.~(\ref{SYMPI}) are in order here.
1)~Since ${\bar m_q}\<\Omega|A_0^1|\pi^1({\bf 0})\>|_{\bar m_q}^{\rm cont}$ is a
quantity of order $\bar m_\pi^3$, the first term is completely immaterial to
the present analysis. 2)~The factor $\bar m_\pi$ in front of the
second term comes from the time derivative in eq.~(\ref{MIX}). 3)~The factor
$2\bar m_\pi$ in the third term arises from the chain of relations
$\<\pi^0({\bf 0})|Q_V^2|\pi^1({\bf q})\>|_{\bar m_q}^{\rm cont}=
\<\pi^0({\bf 0})|\pi^0({\bf q})\>|_{\bar m_q}^{\rm cont}=
(2\pi)^3 2\bar m_\pi|_{\bar m_q}^{\rm cont}\delta^3({\bf q})$. 
4) Concerning the factors $C_L$, $C_{SL}$ and $C_{SSL}$, 
they differ from unit by three different types of lattice artifacts.
i) The terms with $k=\ell=0$ in $C_L$ represent the the leading 
``IR divergent'' corrections contributing to the correlator 
$a^3\sum_{\bf x}\< {\cal Q}(x,0)\>|^{L}_{\bar m_q}$.
ii) Sub-leading ``IR divergent'' corrections are generated by insertions of 
$\int d^4y{\cal L}_{\rm even}(y)$ and/or pairs of $\int d^4y{\cal L}_{\rm odd}(y)$ 
factors. They contribute extra $a^2/\bar m_\pi^2$ powers. 
iii) Finally there are ``IR finite'' corrections 
stemming, among others, from contributions of
intermediate states other than zero three-momentum pions. 

With all these premises, in order to fix the critical mass so as to have in 
the continuum (see eq.~(\ref{EFFCOND}) and the definition~(\ref{METREO})) 
\beq\lim_{\bar{m}_q\to 0^+}\xi_\pi(\bar m_q)=0\, , \label{CONDODD}\eeq
one may think of proceeding on the lattice in the following way. 

Given a first estimate of the critical mass, say $M^e_{\rm cr}$, 
consider the lattice action~(\ref{MTMA}) where the critical mass has 
been momentarily replaced by the expression $M^e_{\rm cr}+\delta\tilde m$.
In order to implement eq.~(\ref{CONDODD}) we must compute 
the lattice quantity in the l.h.s.\ of eq.~(\ref{SYMPI}) and identify 
the optimal value of the critical mass, $M_{\rm cr}^{opt}$, as the 
limiting value of $M^e_{\rm cr}+\delta\tilde m$ at which the O($a^0$) 
quantity (see eqs.~(\ref{LATCOR}) and~(\ref{SYMPI}))
\beq 
A(\bar m_q,M^e_{\rm cr}+\delta\tilde m;t)\equiv
\Big{[}\frac{\bar m_\pi^2}{a} \, e^{\bar m_\pi t} \, 
a^3\sum_{\bf x}\;\<{\cal Q}(x,0)\>\Big{]}^{L}_{\bar m_q} \, , 
\qquad t \gg \frac{1}{\Delta m} 
\label{QA}
\eeq
vanishes as $\bar m_q$ is extrapolated to smaller and smaller values
from the region where $\bar m_q>a$.

Numerically there can be various ways to do this. One possible strategy 
is to start from a value, $\bar{m}^{(1)}_q$, of the quark mass such 
that the order of magnitude inequality $(\bar{m}^{(1)}_\pi)^2>a$ holds.
A first determination, $\delta\tilde m^{(1)}$, of $\delta \tilde m$ 
can be obtained by enforcing at large $t$ the condition 
\beq
A^{(1)}\equiv A(\bar m_q^{(1)},M^e_{\rm cr}+\delta\tilde m^{(1)};t)=0
\qquad t \gg \frac{1}{\Delta m} \, ,
\label{LATCOND1}
\eeq
which in turn yields a first estimate, $\xi^{(1)}_\pi$, of the matrix element 
in~(\ref{CONDODD}). Solving iteratively the non-linear equation~(\ref{LATCOND1}), 
one can write from eqs.~(\ref{SYMPI}) and~(\ref{SYMPICX})
\beq
\xi^{(1)}_\pi = -a(\bar{m}^{(1)}_\pi)^2\eta 
\<\Omega|P^1|\pi^1\!({\bf 0})\>+
\delta \xi^{(1)}_\pi\, , 
\label{FIRES}
\eeq
where the first term is the solution of the linearized form of~(\ref{LATCOND1}) 
obtained by setting to unit the factors $C_{L}$, $C_{SL}$ and $C_{SSL}$ 
in eq.~(\ref{SYMPI}). The second term, $\delta \xi^{(1)}_\pi$, is the correction 
due to all the other terms and, in particular, to the higher powers of $\xi_\pi$. 
To estimate the magnitude of $\delta \xi^{(1)}_\pi$ we plug back the 
ansatz~(\ref{FIRES}) in~(\ref{LATCOND1}). The structure of the result is
\beq
\delta \xi^{(1)}_\pi \, =  \, 
a^3 \sum_{k=0}^{\infty} h_{\; k} 
\Big{(} \frac{ a^2 }{ ( \bar{m}^{(1)}_\pi )^2 }
\Big{)}^{k}  \Big{[}1 
+ {\rm O}( (\bar{m}^{(1)}_\pi)^2 ) 
+ {\rm O}(a^2)\Big{]}
\, . \label{CORRFIRES}
\eeq
where the leading O($a^3$) corrections can be seen to arise from  
the terms with $k=1, j=\ell=0$ in eq.~(\ref{SYMPICX})~\footnote{Actually 
only terms from $C_{L}$ and $C_{SL}$ are important to this order. Terms 
coming from $C_{SSL}$ are, in fact, negligible because of the extra $\bar m_q$ 
factor in front of the corresponding matrix element.}.
The important point about eq.~(\ref{CORRFIRES}) is that 
$\delta \xi^{(1)}_\pi$ can be considered small under 
a condition, $\bar m_q^{(1)}>a^2$, which is weaker than the one 
($\bar{m}^{(1)}_q > a$) we have been using in establishing this result. 

At this point one continues by lowering the quark mass to 
$\bar{m}^{(2)}_q<\bar{m}^{(1)}_q$, seeking for the new value, 
$\delta\tilde m^{(2)}$, of the mass shift which makes $A^{(2)}$ vanishing. 
The search must be performed in the neighborhoods of $\delta\tilde m^{(1)}$ 
to be sure that one remains in a region of the 
($\bar{m}_q,M^e_{\rm cr}+\delta\tilde m$)-plane, where 
$\xi_\pi/\bar m_\pi^2$ is small, so that higher powers of the ratio 
$\xi_\pi/\bar{m}_\pi^2$ are even smaller. Proceeding in this way the 
convergence of the Symanzik expansion~(\ref{SYMPI}) is not put in 
danger as $\bar{m}_q$ is decreased. Rather, the initial convergence
bound $\bar m_q>a$ is progressively weakened towards $\bar m_q>a^2$,
as signaled by eq.~(\ref{CORRFIRES}).

A sequence, $\delta\tilde m^{(i)}$, of mass shifts is thus determined. 
If desired, these values can be numerically extrapolated to $\bar m_q\to 0$. 
The limiting value, $\delta\tilde m^{(\infty)}$, obtained in this way will 
allow to identify the ``optimal'' critical mass as the quantity 
\beq
M^{opt}_{\rm cr}=M^e_{\rm cr}+\delta\tilde m^{(\infty)} \, .
\label{MCROPT} \eeq  

Notice that in all this procedure no lattice data points are employed 
where the bound $\bar m_q>a^2$ is violated. This caution is necessary because 
for $\bar m_q \leq a^2$ large cutoff effects, which are hinted at 
by uncanceled non-leading ``IR divergent'' terms in the Symanzik 
expansion, cannot in general be excluded and reliable simulations
may even be impossible because of metastabilities, if e.g.\ the peculiar
lattice phase structure known as the ``Sharpe-Singleton scenario'' 
\cite{PS} is realized.

The method discussed above may seem unpractical in view of the fact 
that, especially for unquenched simulations, producing data at several 
values of the bare quark mass, as it is necessary to do if one wants to 
extrapolate to $\bar m_q\to 0$, is computationally rather demanding. 
We immediately notice, however, that to all practical purposes (see the 
detailed argument given in Sect.~\ref{sec:PSIM} below) 
one can avoid such an extrapolation and just work with the 
``quasi-optimal'' critical mass   
\beq
M^{qopt}_{\rm cr}(\bar{m}_q^{\rm min}) \equiv
M^e_{\rm cr}+\delta\tilde m|_{\bar{m}_q^{\rm min}} \, ,
\label{MCRQOPT}
\eeq
where $\bar{m}_q^{\rm min} > a^2$ is the smallest of the 
bare quark masses of interest at the lattice spacing one is working and 
$\delta\tilde m|_{\bar{m}_q^{\rm min}}$ is the solution of the equation 
$A(\bar m_q^{\rm min},M^e_{\rm cr}+\delta\tilde m|_{\bar{m}_q^{\rm min}};t)=0$. 

The idea of estimating the critical mass at the smallest available 
value of $m_q$ was already put forward in the analysis of tm-LQCD 
performed in $\chi$PT in refs.~\cite{SHWUNEW,AB} and directly used in 
the unquenched simulations of ref.~\cite{KARETAL}. 
In quenched simulations slightly different versions of the strategy 
described above have been implemented in order to evaluate the 
(quasi-)optimal value of the critical mass. They all turned out to be 
quite effective in reducing cutoff artifacts~\cite{CAN,XLFNEW,XLFSCAL}.

\subsubsection{Left-over ``IR divergent'' cutoff effects}
\label{sec:PSIM}

To complete our analysis we have to determine what is the order of 
magnitude of the left-over discretization errors that will affect 
simulations carried out at non-vanishing $m_q$, when either 
i) eq.~(\ref{MCROPT}) or ii) eq.~(\ref{MCRQOPT}) for 
the critical mass is inserted in the Mtm-LQCD action~(\ref{MTMA}). 
We will see that in both cases the situation will be very much like 
the one we encountered in Sect.~\ref{sec:IASWT}, where we discussed the 
case in which the clover term was added to the Mtm-LQCD action. 

i) Let us denote by the superscript ``\,$^{opt}$\,'' lattice 
quantities computed using in the fermionic action the optimal critical 
mass, $M^{opt}_{\rm cr}$ (eq.~(\ref{MCROPT})).
The formerly leading $2k$-fold pion pole contribution~(\ref{ZLCP}) 
now have an expression where $\xi_\pi(m_q)$ is systematically
replaced by $\xi_\pi^{opt}(m_q)$, which represents the value of the 
continuum matrix element  
$|\<\Omega|{\cal L}_{\rm odd}^{opt}|\pi^0({\bf 0})\>|$
at the current value, $m_q$, of the quark mass,
with ${\cal L}_{\rm odd}^{opt}$ the parity odd part of the Symanzik 
LEEA when the optimal critical mass is employed. 

Although non-zero at non-vanishing quark mass, we want to show 
that $\xi_\pi^{opt}(m_q)/m_\pi^2={\mbox O}(a)$.  
To this end we notice that one can write
\beq
\xi_\pi^{opt}(m_q)=\xi_\pi^{opt}(0)+
\frac{\partial\xi_\pi^{opt}(m_q)}{\partial m_q}
\Big{|}_{m_q=0}m_q+\ldots\, ,
\label{DERIV}\eeq
where from eqs.~(\ref{FIRES}) and~(\ref{CORRFIRES}) one has  
$\xi_\pi^{opt}(0)={\mbox O}(a^3)$. This result comes from having
extrapolated to $\bar m_q=0$ from the region $\bar m_q>a^2$, where 
$\delta\xi_\pi={\mbox O}(a^3)$. Since 
$\partial\xi_\pi^{opt}(m_q)/\partial m_q|_{m_q=0}\sim{\mbox O}(a)$, 
we conclude that 
$\xi_\pi^{opt}(m_q)$ is reduced to a mere order $a m_q$ quantity,  
implying, as announced, $\xi_\pi^{opt}(m_q)/m_\pi^2={\mbox O}(a)$.

ii) If we set instead $M^e_{\rm cr}=M^{qopt}_{\rm cr}(\bar{m}_q^{\rm min})$ 
(eq.~(\ref{MCRQOPT})) in the action~(\ref{MTMA}), we get (notation should be 
self-explanatory) 
\beq
\xi_\pi^{qopt}(m_q)=\xi_\pi^{qopt}(\bar m_q^{\rm min})+
\frac{\partial\xi_\pi^{qopt}(m_q)}{\partial m_q}
\Big{|}_{\bar m_q^{\rm min}}(m_q-\bar m_q^{\rm min})+\ldots\, ,
\label{DERIV1}\eeq
where from eqs.~(\ref{FIRES}) and~(\ref{CORRFIRES}) one finds  
$\xi_\pi^{opt}(\bar m_q^{\rm min})={\mbox O}(a\bar m_q^{\rm min})$,
provided $\bar m_q^{\rm min}>a^2$. From 
$\partial\xi_\pi^{qopt}(m_q)/\partial m_q |_{\bar m_q^{\rm min}}\sim{\mbox O}(a)$ 
one obtains for any $m_q\geq\bar{m}_q^{\rm min}$ the estimate 
$\xi_\pi^{qopt}(m_q)/m_\pi^2={\mbox O}(a,a\bar{m}_q^{\rm min}/m_q)={\mbox O}(a)$.

The conclusion of this analysis is that both with the optimal
as well as the quasi-optimal critical mass, 
the formerly leading ``IR divergent'' cutoff discretization 
errors are reduced to finite O($a^{2k}$) contributions. 

This does not mean that all 
the non-leading ``IR divergent'' cutoff effects, which are
of order $a^{2k}/(m_\pi^2)^{h}$, $2k>h\geq 1$ ($k,h$ integers),
have disappeared from correlators. Actually by a non-trivial diagrammatic 
analysis, based on the structure of the non-leading ``IR divergent'' cutoff 
contributions, one can prove that the most ``IR divergent'' lattice artifacts
left behind after using either the optimal or the quasi-optimal 
definition of the critical mass (eq.~(\ref{MCROPT}) or
eq.~(\ref{MCRQOPT})) are reduced down to order $a^2(a^2/m_\pi^2)^{k-1}$ 
($k\geq 1$) effects at worse, just like in the case the clover term is 
employed. Notice, however, that in the case where the 
quasi-optimal critical mass is adopted this result holds only for 
$m_q\geq\bar{m}_q^{\rm min}$.

\section{The lattice GMOR relation}
\label{sec:GOR}

In the previous sections we have shown how the leading ``IR divergent'' 
discretization effects can all be eliminated from the Symanzik expansion of 
lattice correlators. The left-over ``IR divergent'' terms of the expansion 
turn out to have the structure $a^{2}\sum_{\ell\geq 1}c_\ell(a^2/m_q)^\ell$. 
The convergence of this series sets the order of magnitude inequality 
$a^2<m_\pi^2\sim m_q$ from which one should determine the minimal 
value of the quark mass that (at fixed value of $a$) can be safely simulated 
before possibly encountering large discretization errors. 

A workable way to numerically estimate this minimal value can be obtained by 
considering the behaviour of the charged pion mass as a function of $m_q$. 
It turns out, in fact, that in Mtm-LQCD there are Ward-Takahashi identities 
(WTI's) which take exactly the form they have in the formal continuum theory. 
{}From them a lattice GMOR relation can be derived.

To see how this works we recall that in Mtm-LQCD the 1-point split axial 
currents, $\hat{A}_\mu^b$, with iso-spin index $b=1,2$ are exactly conserved 
in the chiral limit $m_q=0$~\cite{TM,FR1}. This implies the validity of the 
lattice WTI's
\beq
\<\Big{[}\partial_\mu^* \hat A_\mu^\pm(x)-2m_q P^\pm(x)\Big{]} 
P^\mp(0)\>\Big{|}^L_{m_q}= 
\< S^0(0)\>\Big{|}^L_{m_q} \delta_{x,0}\, ,
\label{PCAC}\eeq 
\beq
\hat{A}_\mu^\pm=\hat{A}_\mu^1\pm i\hat{A}_\mu^2\, ,\qquad 
P^\pm=\bar\psi^L\gamma_5\frac{\tau_1\pm i\tau_2}{2}\psi^L\, ,\qquad 
S^0 =\bar\psi^L\psi^L\, .
\label{P1S0}\eeq
After integration over space-time, one gets for any $m_q\neq 0$
\beq
2m_q\, a^4\sum_{x}\< P^\pm(x)P^\mp(0)\>\Big{|}^L_{m_q}=
-\< S^0(0)\>\Big{|}^L_{m_q} \, .
\label{PCACI}\eeq
Although, as the WTI~(\ref{PCACI}) itself shows, there is no mixing between 
$S^0$ and the identity operator with a cubically (or a linearly) divergent 
coefficient, there is still room for a quadratically divergent term proportional 
to $m_q$. Indeed, the l.h.s.\ of eq.~(\ref{PCACI}) is equal to the piece 
where intermediate states are inserted plus a divergent 
contribution of the kind $m_q/a^2$ coming from the (integrated) short-distance 
singularity of the correlator $\<P^\pm(x)P^\mp(0)\>$ at $x~\!=~\!0$. 
This term should be brought to the r.h.s.\ of the equation, thus
leaving finite (subtracted) expressions in both sides of the resulting equation.

We can now repeat on the lattice the argument that in the continuum
leads to the classical GMOR relation, if we assume that spontaneous 
chiral symmetry breaking occurs in the limiting continuum theory, 
i.e.\ if we assume that  
\beq
\Sigma \equiv - \lim_{m_q\to 0}\<\Omega|S^0|\Omega\>\Big{|}^{\rm cont}_{m_q}\neq 0\, .
\label{SCSB}\eeq
As in the formal continuum theory, we insert a complete set of states 
in the subtracted correlator
in the l.h.s.\ of eq.~(\ref{PCACI}). We obtain in this way
\beq
m^2_{\pi^\pm}\Big{|}^L_{m_q}=
{2m_q}\left.\frac{|\<\Omega|P^\pm|\pi^\pm\>|^2}{[-\<\Omega|S_{\rm sub}^0|\Omega\>]}
\right|^L_{m_q} + \ldots\, ,\label{PCACIR}\eeq
where we have explicitely written down only the contribution coming from the pion 
pole. Dots are terms due to the intermediate states that stay massive 
as $m_q \to 0$ as well as terms vanishing with $m_q$ faster than linearly
(the latter include in particular O($a^2m_q^2$) terms stemming from cutoff
effects in the sum over $x_0$~\footnote{No O($a$) terms can arise from
the sum over $x_0$ in the l.h.s.\ of eq.~(\ref{PCACI}),
as in Mtm-LQCD symmetry arguments ensure that only O($a^2$) discretization errors 
can affect expectation values of parity even operators.}). 
The subscript ``$_{\rm sub}$'' is to remind us that it is the properly 
subtracted chiral condensate (or, more precisely, the vacuum expectation
value of the properly subtracted iso-singlet scalar density operator)  
that enters this equation. Notice that, as expected, the r.h.s.\ of eq.~(\ref{PCACIR}) 
is a finite renormalization group invariant quantity in the limit $a\to 0$.

Once the leading ``IR divergent'' cutoff effects have been canceled out,
the use of the Symanzik expansion in the r.h.s.\ of eq.~(\ref{PCACIR}) 
yields the formula 
\beqn
\hspace{-.8cm}&&m^2_{\pi^\pm}\Big{|}^L_{m_q}=
{2m_q}\left.\frac{|\<\Omega|P^\pm|\pi^\pm\>|^2}{\Sigma}\right|^{\rm cont}_{m_q=0}
\Big{[}1+a^2\sum_{\ell\geq 0}b_\ell
\Big{(}\frac{a^2}{m_q}\Big{)}^\ell \Big{]}+\ldots
\, ,\label{PCASYM}\eeqn
where dots denote less dangerous ``IR divergent'' lattice artifacts 
compared to those explicitely shown as well as contributions of higher 
order in $m_q$. In getting eq.~(\ref{PCASYM}) we have used the fact 
that the continuum limit of $-\<\Omega|S_{\rm sub}^0|\Omega\>^L$  
at vanishing quark mass is $\Sigma\neq 0$ (see eq.~(\ref{SCSB})).

{}From the above analysis it follows that, in the region where 
the series in eq.~(\ref{PCASYM}) converges (i.e.\ at least where 
the order of magnitude inequality $m_\pi^2\sim m_q >a^2$ is 
satisfied), the squared mass of the charged lattice pion is linear in 
$m_q$ (up to small O($a^4$) and O($m_q^2$) effects). Thus, vice-versa,
we can imagine to use deviations from the established linear behaviour 
possibly seen at small $m_q$ as a workable criterion to determine the 
minimal value of $m_q$ at which simulations can be performed before 
being set-off by discretization effects. 

\section{Hadron masses and pion decay constant}
\label{sec:ART}

We wish to discuss in this section some peculiar issues concerning 
the magnitude of the O($a^2$) discretization artifacts affecting 
hadronic energies (in particular masses) and the pion decay constant.

\subsection{Hadron energies}
\label{sec:PM}

As a consequence of the special form of the diagrams contributing to 
energies (self-energy diagrams), one can get convinced that the latter 
are less ``IR divergent'' than the correlators from which they can be 
extracted, in the sense that at fixed order in $a$, the most ``IR divergent'' 
lattice corrections to their continuum limit contain one overall 
factor $1/m_\pi^2$ less than the ``IR divergent'' cutoff 
effects appearing in correlators. The reason is that at least one 
among the inserted $\int {\cal L}_{\rm odd}$ operators gets absorbed 
in a multi-particle matrix element, with the consequence that it is 
not anymore available for producing a pion pole 

An explicit calculation shows that for the difference between lattice 
and continuum energy of the hadron $\alpha_n$ 
\beq
\Delta E_{\alpha_n}({\bf q}) \equiv \frac{1}{2}[E^L_{\alpha_n}({\bf q})
+ E^L_{\alpha_n}({\bf -q}) ] \, - \, E^{\rm cont}_{\alpha_n}({\bf q})\, ,
\label{DE2} \eeq
one gets at order $a^2$ the estimate 
\beqn
\hspace{-.8cm}&&\Delta E_{\alpha_n}({\bf q})\Big{|}_{a^2}\propto\nn\\
\hspace{-.8cm}&&\propto \left.
\frac{a^2}{m_\pi^2}{\rm Re}\frac{\<\Omega| \ell_5 | \pi^0({\bf 0}) \>
\<\pi^0({\bf 0})\alpha_n({\bf q})| \ell_5 |\alpha_n({\bf q})\>_{\rm conn} }
{2 E_{\alpha_n}({\bf q}) } 
\left[  1 + {\rm O}(m_\pi^2) \right]
\right|_{m_q}^{\rm cont}\, ,
\label{DE2LEAD}\eeqn
where the subscript $_{\rm conn}$ denotes 
the completely connected part of the matrix element
$\<\pi^0({\bf 0})\alpha_n({\bf q})| \ell_5 |\alpha_n({\bf q})\>$. 
In eq.~(\ref{DE2LEAD}) the ``IR divergent'' O($a^2/m_\pi^2$) piece 
comes from the continuum correlator with two 
${\cal L}_{\rm odd} = a\ell_5 + {\rm O}(a^3)$ insertions. The latter
gives also rise to O($a^2$) ``IR finite'' corrections. A further 
O($a^2$) ``IR finite'' correction comes from a single insertion of
${\cal L}_{\rm even} = a^2 \ell_6 + {\rm O}(a^4)$ and contributes
a term proportional to 
$a^2 \<\alpha_n({\bf q})|\ell_6|\alpha_n({\bf q})\> /2E_{\alpha_n}({\bf q})$. 

It should be noted that the ``IR divergent'' lattice artifact in 
$\Delta E_{\alpha_n}|_{a^2}$ is reduced to an ``IR finite'' correction 
after anyone of the two ``cures'' described in Sect.~\ref{sec:KLL}. 

Specializing the formula~(\ref{DE2LEAD}) to the case of pions, one 
obtains the interesting result that the difference between charged 
and neutral pion (square) masses is a finite O($a^2$) quantity even 
if the critical mass has not been set to its ``optimal'' value or the 
clover term has not been introduced. The reason is that the 
leading ``IR divergent'' contributions shown in eq.~(\ref{DE2LEAD}) 
are equal for all pions (as one can prove by standard soft pion 
theorems~\cite{SPT}), hence cancel in the (square) mass difference. 
This conclusion is in agreement with results from $\chi$PT~\cite{PS}. 

\subsection{Pion decay constant}
\label{sec:PDC}

Data on a (quenched) computation of $f_\pi$, carried out 
in Mtm-LQCD using the value of the critical mass obtained from
the vanishing of the pion mass, show cutoff effects that are 
seen to deviate at small bare quark masses from the straight line 
extrapolation drawn from large masses~\cite{KSUW}.
This behaviour (called ``bending phenomenon'' in ref.~\cite{KSUW}) 
sets in at values of $m_q$ around $a \Lambda_{\rm QCD}^2$, 
$\Lambda_{\rm QCD}\sim 200$~MeV. Furthermore the detailed recent scaling 
test of~\cite{XLFSCAL} indicates that the ``bending phenomenon'' is 
an O($a^2$) deviation with a magnitude which increases 
as $m_{\pi^\pm}^2$ is lowered. 

In the works of refs.~\cite{KSUW,CAN,XLFNEW,XLFSCAL} the lattice 
pion decay constant, $f_\pi^L$, was extracted from the formula~\cite{FR1}
\beq
f_\pi^L(m_{\pi^\pm}^2)=2m_q\frac{\<\Omega|P^\pm|\pi^\mp\>}
{m_{\pi^\pm}^2}\Big{|}^L_{m_q}\, ,\label{FPIL}\eeq
from which it is seen that $f_\pi^L$ is a ratio of two lattice quantities.

It should be noted that in general the matrix element 
$\<\Omega|P^\pm|\pi^\mp\>|_{m_q}^L$ in the numerator is affected by leading 
``IR divergent'' cutoff effects which however, like in the case of hadron
masses, are multiplied by an extra factor of $m_\pi^2\sim m_q$. 
This peculiar property can be traced back to the invariance of the 
correlator $\<P^\pm(x) P^\mp(y)\>|_{m_q}^L$ (from which the matrix 
element $\<\Omega|P^\pm|\pi^\mp\>|_{m_q}^L$ is extracted) 
under axial rotations around  the third iso-spin direction. 
At the price of shifting the bare quark mass from $m_q$ to 
$\widetilde{m}_q = m_q + {\rm O}(a^2/m_q)$ one can, in fact, 
always think of having performed the axial rotation with an angle 
such to bring the critical mass to its optimal value. 
A Symanzik expansion of $\<P^\pm(x) P^\mp(y)\>|_{\widetilde{m}_q}^L$ 
evaluated with the optimal critical mass then shows 
that the leading ``IR divergent'' cutoff effects (which
from this viewpoint arise owing to $\widetilde{m}_q\neq m_q$ only) 
are softened by the extra multiplicative $m_\pi^2$ factor. 
Similarly it follows from the discussion of Sect.~\ref{sec:PM} 
(which is consistent with the previous argument) that also 
$m_{\pi^\pm}^2|^L_{m_q}$ is affected by leading ``IR divergent'' cutoff 
artifacts softened by an overall $m_\pi^2$ factor. The latter, however, 
drops when the relative lattice correction is considered. 

Inserting in eq.~(\ref{FPIL}) the behaviour of numerator and denominator 
we have just discussed, we find that $f_\pi^L$ is in general affected 
by ``IR divergent'' {\em relative} corrections that are fully leading 
in our nomenclature, i.e.\ of the kind $(a/m_\pi^2)^{2k}$, $k\geq 1$. 
Of course, if the appropriate clover term or the ``optimal'' critical mass 
is introduced in the action, these ``IR divergent'' relative lattice 
artifacts are reduced to $a^{2}(a^2/m_\pi^2)^{k-1}$, $k\geq 1$ effects.

To be concrete at order $a^2$ one finds that in $f_\pi^L$ (see 
eq.~(\ref{FPIL})) relative corrections of the type $a^2/m_q^2$ are 
present, which come only from $(m_{\pi^\pm}^2/m_q)|^L_{m_q}$. 
Less important corrections of the type $a^2/m_q$ and $a^2$ come 
instead from both $\<\Omega|P^\pm|\pi^\mp\>|_{m_q}^L$ and 
$(m_{\pi^\pm}^2/m_q)|^L_{m_q}$. After the clover improvement or the 
``optimal'' critical mass cure, the relative cutoff effects in $f_\pi^L$ 
get reduced to plain $a^2$ terms. Results consistent with these features
have also been derived in $\chi$PT studies of 
Mtm-LQCD~\cite{SHWUNEW,BAER,SH05} at (next-to) leading order. 

A beautiful confirmation of the validity of the analysis presented in this 
section comes from the fact that, when the critical mass is set at its 
``optimal'' value (yielding $\xi_\pi^{opt}(m_q) = {\rm O}(am_\pi^2)$), 
no ``bending phenomenon'' is anymore visible 
in the $f_\pi^L$ data, as demonstrated by the results of 
ref.~\cite{CAN,XLFNEW,XLFSCAL} (see also~\cite{ASH}).

\section{Conclusions}
\label{sec:CONC}

We have shown in this paper that lattice correlators in Mtm-LQCD are  
affected by discretization artifacts that tend to become large when the 
quark mass is decreased. Cutoff effects of this kind can be appropriately 
described in terms of the Symanzik LEEA of the lattice theory, and turn 
out to arise from multiple pion poles associated with the insertions of 
the parity odd (and iso-spin non-invariant) piece of the effective action, 
$\int \!d^4y\,{\cal{L}}_{\rm odd}(y)$.

At order $a^{2k}$ the leading ``IR divergent'' cutoff effects (those with
$h=2k$ in eq.~(\ref{ORD})) are associated to $2k$-fold pion poles.
Since their residues contain $2k$ factors of 
the matrix element $\xi_\pi=|\<\Omega|{\cal L}_{\rm odd}|\pi^0({\bf 0})\>|=
|a\<\Omega|{\ell}_{5}|\pi^0({\bf 0})\>|+{\rm O}(a^3)$, all these dangerous 
contributions are eliminated from lattice correlators if we can let 
$\xi_\pi$ vanish sufficiently fast as the pion mass goes to zero. 
This can indeed be achieved if Mtm-LQCD is improved {\it \`a la} 
Symanzik by the inclusion of the lattice clover term in the action or by 
appropriately tuning the critical mass parameter to what we have called 
its ``optimal'' value. 

In both cases the left-over ``IR divergent'' cutoff effects are 
at worse of order $a^2(a^2/m_\pi^2)^{k-1}$, $k\geq 1$. 

The discussion we presented in this work does not make use of chiral 
perturbation theory and treats $m_q$ as a fixed quantity in the limit 
in which $a\to 0$. By using the notion of Symanzik LEEA, one is able 
to obtain results to all orders in $a$. 
It must be recalled, however, that, as recently pointed 
out in ref.~\cite{SH05}, a $\chi$PT analysis of (M)tm-LQCD 
shows that it is possible to reabsorb (at any given order in the 
chiral power counting of choice) a whole tower of cutoff effects in a 
shift of the chiral lattice vacuum, thus extending the radius of 
convergence of the associated expansion~\footnote{We are indebted to 
S. Sharpe and M. Golterman for valuable discussions on this point.}. 

Our detailed findings are in agreement with the results of 
refs.~\cite{AB,SHWUNEW,SH05} concerning the quark mass dependence 
of lattice artifacts and the existence of an ``optimal'' choice for 
the critical mass yielding a complete cancellation of the leading 
``IR divergent'' cutoff effects. In particular, we agree with the authors 
of ref.~\cite{SHWUNEW} that, once the ``optimal'' value of the critical 
mass is employed in Mtm-LQCD simulations (or the clover 
term is introduced), the tight (order of magnitude) limitation $m_q>a$ 
on the quark mass is substantially weakened and brought down 
to the much more favourable bound $m_q > a^2$, beyond which large 
mass-dependent cutoff effects may show up in simulations at fixed $a$.

%%%%%%%%%%%%%%%%%%%%%%%%%%%%%%%%%%%%%

We conclude this paper with a remark which explains how the optimal
critical mass prescription can be viewed as a peculiar mapping
between bare and renormalized parameters and may be useful in practice 
to correct for a not optimal choice of the critical mass. Indeed, the 
fact that it is possible to get rid of all the leading ``IR divergent''
cutoff effects by working at an optimal value of the critical
mass implies that, even if simulations are carried out with
a non-optimal choice of $M_{\rm cr}$, one can still effectively
eliminate all the dangerous $(a/m_q)^{2k}$ ($k\geq 1$) terms by
means of a slightly more elaborate analysis of the available simulation
data. This analysis consists in using an appropriately redefined
expression of the renormalized quark mass (by O($a^2$) terms), accompanied
by a small (O($a$)) chiral rotation in the third iso-spin direction
of all quantities that have non-trivial chiral transformation
properties~\footnote{This is the kind of procedure we have alluded
to in sect.~\ref{sec:PDC}, below eq.~(\ref{FPIL}). Detailed formulae
for this analysis can be obtained on the basis of the results
in ref.~\cite{FR1} about the relation between bare and renormalized
mass parameters and currents (or more general operators) in tmLQCD.}.
By construction the lattice correlators that are obtained in this way
will have a Symanzik expansion where the leading "IR-divergent" cutoff
effects are just absent, while automatic O($a$) improvement is
obviously preserved (the deviations from the initial simulation values
of the critical mass and twist angle are in fact O($a$) quantities).
However this analysis requires some knowledge (or a simultaneous
determination, which can be obtained by working at several values of
$(M_0,m_q)$) of scale-independent combinations of the renormalization
constants of operators belonging to the same chiral multiplet, such as
$Z_P/Z_{S^O}=Z_P Z_m$ or $Z_V/Z_A$.

Indeed the prescription we have described is the microscopic counter-part
of what is suggested in ref.~\cite{SHWUNEW} at the level of the
effective chiral theory in order to reabsorb the contributions of these
lattice artifacts via a proper redefinition of the vacuum state.

Being based on the assumption that an acceptable (i.e.\ O($a$) accurate)
estimate of the critical mass is already known, this remark is of no use
to obtain such an estimate (the argument would become circuitous).
It should  also be observed that the effort of determining by how much the
renormalized quark mass have to be shifted and chiral operators rotated is
essentially equivalent (modulo the knowledge of $Z_P/Z_{S^O}$ and analogous 
renormalization constant ratios) to enforce the optimality condition in
the way we have described it in sect.~\ref{sec:OCCM}.

%%%%%%%%%%%%%%%%%%%%%%%%%%%%%%%%%%

\vskip .2cm
{\bf Acknowledgments -- } We are grateful to the anonymous referee
for his valuable observations.
We wish to thank K. Jansen, M. L\"uscher, S. Sharpe and M. Testa  
for useful discussions and V. Lubicz and M. Golterman for correspondence. 
We are grateful to K. Jansen, M. Papinutto, A. Shindler, C. Urbach 
and I. Wetzorke for information about the results of ref.~\cite{XLFNEW} 
prior to publication. One of us (GCR) wishes to thank the Humboldt 
Foundation for financial support.

\appendix
\renewcommand{\thesection}{Appendix~A}
\section*{Appendix - O($a^{2k+1}$) improvement of Mtm-LQCD}
\renewcommand{\thesection}{A}
\label{sec:PARI}

We prove in this Appendix the absence of all the terms of order $a^{2k+1}$,
$k\geq 0$, in the Symanzik expansion of the expectation value of parity-even
(multi-local) gauge-invariant and multiplicatively renormalizable operators
in Mtm-LQCD. Here to be precise by ``parity-even'' we mean an operator 
which goes exactly into itself under parity. This result follows from the 
symmetry $P\times {\cal{D}}_{d}\times (m_q\to -m_q)$, enjoyed by the lattice 
action, where ($x_P=(-{\bf{x}},t)$)  
\begin{equation}  
{\cal{P}}:\left \{\begin{array}{lll} 
U_0(x)&\rightarrow &U_0(x_P)\, ,\\
U_k(x)&\rightarrow &U_k^{\dagger}(x_P-a\hat{k})\, ,\qquad k=1,2,3\\
\psi(x)&\rightarrow &\gamma_0 \psi(x_P)\\
\bar{\psi}(x)& \rightarrow &\bar{\psi}(x_P)\gamma_0 
\end{array}\right . \label{PAROP}  \end{equation} 
is the physical parity of the theory,
and 
\begin{equation}    
{\cal{D}}_d : \left \{\begin{array}{lll}     
U_\mu(x)&\rightarrow & U_\mu^\dagger(-x-a\hat\mu) \, ,\qquad \mu=0,1,2,3\\
\psi(x)&\rightarrow & e^{3i\pi/2} \psi(-x)  \\
\bar{\psi}(x)&\rightarrow & e^{3i\pi/2} \bar{\psi}(-x)  
\end{array}\right . \label{FIELDT} \end{equation} 

The proof of O($a^{2k+1}$) improvement of Mtm-LQCD follows immediately
from the observation that necessarily the Symanzik LEEA which 
describes the lattice artifacts of Mtm-LQCD is invariant under (the 
continuum version of) the transformation 
$P\times {\cal{D}}_{d}\times (m_q\to -m_q)$. This invariance implies, 
in particular, that all the terms of order $a^{2k+1}$, $k\geq 0$, 
in~(\ref{SLEEA}) are odd under parity. In fact, 
${\cal{D}}_d\times (m_q\to -m_q)$ counts the parity of the 
overall dimension, $d$, of any product of elementary fields and mass factors, 
by multiplying it by the phase factor $\exp(i\pi d)$ (besides inverting its 
space-time argument).
As a result, all the continuum correlators in the Symanzik expansion 
that are multiplied by an odd power of $a$ necessarily 
correspond to expectation values of parity odd operators. 
Since parity is an exact continuum symmetry, they all vanish.

Few comments are in order here.

1) It is interesting to observe that automatic O($a^{2k+1}$), $k\geq 0$, 
improvement (in the Symanzik sense) is a very robust property of Mtm-LQCD. 
Being based on straight symmetry arguments, it is independent of any 
consideration on possible phase transition scenarios~\cite{AB,PS}.
In fact, for any value of $m_q$, sufficiently close to the continuum limit 
the chiral phase of the vacuum will be finally determined by the quark 
mass term in the action and not by the Wilson term.

2) At maximal twist the $r$-parity properties of the critical mass are 
immaterial to the above argument for automatic O($a^{2k+1}$), $k\geq 0$, 
improvement. The reason is that in this situation (but not for generic 
values of the twist angle) the Wilson term, and thus necessarily also its 
critical mass counter-term, are odd under parity. This clearly 
remains true for whatever determination of the critical mass 
is taken, independently of its behaviour under $r \to -r$.

On the contrary, at non-maximal twist, the proof of O($a$) improvement 
(via Wilson averaging) requires the critical mass to be an odd function 
of $r$. As explained in the papers of ref.~\cite{FR1}, this condition can 
(and should) always be imposed when determining $M_{\rm cr}(r)$.

3) Finally we note that the conclusions about automatic O($a^{2k+1}$) 
improvement reached in this Appendix can be extended to the case of maximally 
twisted quarks with non-degenerate masses~\cite{FR2}. If the physical basis 
of ref.~\cite{FR2} is employed, a proof completely analogous to that given 
above for mass degenerate quarks can be given relying now on the symmetry 
$P\times{\cal{D}}_{d}\times(m_q\to -m_q)\times(\epsilon_q\to -\epsilon_q)$,
with $\epsilon_q$ the bare mass splitting within the quark pair.

\end{document}